# Tuneable Magneto-Resistance by Severe Plastic Deformation

**Stefan Wurster [1,*], Lukas Weissitsch [1], Martin Stückler [1], Peter Knoll [2], Heinz Krenn [2], Reinhard Pippan [1] and Andrea Bachmaier [1]**

[1] Erich Schmid Institute of Materials Science of the Austrian Academy of Sciences, Jahnstrasse 12, 8700 Leoben, Austria;

[2] Institute of Physics, University of Graz, Universitätsplatz 5, 8010 Graz, Austria;

* Correspondence: stefan.wurster@oeaw.ac.at

Abstract

Bulk metallic samples were synthesized from different binary powder mixtures consisting of elemental Cu, Co, and Fe using severe plastic deformation. Small particles of the ferromagnetic phase originate in the conductive Cu phase, either by incomplete dissolution or by segregation phenomena during the deformation process. These small particles are known to give rise to granular giant magneto-resistance. Taking advantage of the simple production process, it is possible to perform a systematic study on the influence of processing parameters and material compositions on the magneto-resistance. Furthermore, it is feasible to tune the magneto-resistive behavior as a function of the specimens' chemical composition. It was found that specimens of low ferromagnetic content show an almost isotropic drop in resistance in a magnetic field. With increasing ferromagnetic content, percolating ferromagnetic phases cause an anisotropy of the magneto-resistance. By changing the parameters of the high pressure torsion process, i.e., sample size, deformation temperature, and strain rate, it is possible to tailor the magnitude of giant magneto-resistance. A decrease in room temperature resistivity of ~3.5% was found for a bulk specimen containing an approximately equiatomic fraction of Co and Cu.

1. Introduction

The giant magneto-resistance (GMR), independently discovered by the two groups of Fert and Grünberg [1,2] at the end of the 1980s, was first observed for stacks of very thin multilayers of alternating ferromagnetic/antiferromagnetic Fe and Cr. These layers couple magnetically, resulting in a giant decrease of resistivity with an increasingly applied magnetic field. Some years after the discovery of GMR, it was found that this phenomenon is not only restricted to layered systems but can also be found for materials containing dispersed ferromagnetic particles (granules) [3,4]; thus it was labeled as granular GMR.

If grains are not subjected to an external magnetic field, a random orientation of magnetic moments or domains prevails. With an increasing magnetic field, the magnetic domains gradually align by rotating the magnetization, and become aligned parallel to the magnetic field. This results in an overall decrease of resistance. It was shown that the increase in resistance for randomly oriented magnetic particles originates from spin-dependent scattering of conduction electrons at the magnetic-nonmagnetic interfaces [5,6]. Rabedeau et al. [5] found this fact by using small angle X-ray scattering measurements on thin films, making the particle sizes accessible. If the GMR originates from scattering within the ferromagnetic particles, the GMR would weakly depend on the particle size (providing that all the ferromagnetic atoms can be found in the particles). However, as GMR scaled with the inverse of the cluster size (interfaces per volume, $r^2/r^3$) instead, an interfacial spin-dependent scattering was proposed.

Upon applying a magnetic field, the gradual change in the magnetization direction of single domain particles leads to a gradual change of the resistivity and this property can be directly linked to the hysteresis loop of the material. One method to characterize the relationship between the specimens' resistance and the magnetic field is the squared global relative magnetization $\mu(H) = (M(H)/M_S)^2$ [4], which is the ratio of the magnetization $M(H)$ at a certain applied field $H$ and the saturation magnetization $M_S$. The GMR-effect is described in the following way:

$$GMR = \frac{\Delta R}{R} = \frac{R(H) - R(H=0)}{R(H=0)} = A\,\mu(H)^2, \qquad (1)$$

where A determines the effect amplitude and is different for each experimental setup. In some cases, the GMR of Equation (1) is expressed by R (H) in the denominator instead of R (H = 0), or R (H = 0) is replaced by R (H = H$_C$). It is stated [4] that the change in resistance, which is a measure for the GMR-effect, is proportional to A* (M (H) / Ms)$^2$; thus, the strength of the GMR-effect can be quantified by the proportionality factor A. For a magnetron sputtered thin film specimen consisting of 84 at% Cu and 16 at% of Co (Cu84Co16) and a temperature of 5 K, this proportionality factor A was found to be ~0.065 [4]. To allow comparison, all compositions in this work will be given in at%, except stated otherwise.

Research on granular GMR first started with thin films. They were produced using a variety of different techniques, such as magnetron sputtering [3,4,7–11], molecular beam epitaxy [5], ion beam co-sputtering [12], cluster beam deposition [13], thermal evaporation [14], or by electrochemical deposition [15–17]. Later, research was extended to bulk materials and mixed granular materials were produced using different techniques such as mechanical alloying/ball milling [18–22] or melt spinning [9,23–25]. Research focused on a small number of binary, sometimes ternary systems such as CuCo [3–5,9,14,16,17,19,21,23,25,26], AgCo [4,6,7,9,11,12,18], CuFe [13,19], CoFe-Cu [10,20,22], CrFe [8],

and AuCo [9,24]. A common feature of these systems is the small mutual solubility of ferromagnetic and non-magnetic elements, as well as the nonmagnetic phase representing the major phase. Reduced solubility promotes the production of small, finely dispersed ferromagnetic particles within a nonmagnetic metal matrix; either directly during production or after adequate annealing treatments. With increasing ferromagnetic content, a transition towards anisotropic magneto-resistance (AMR) is found [10]. Thomson discovered this anisotropy of ferromagnetic materials in magnetic fields [27], where the resistance for currents parallel and perpendicular to the magnetic field is different. For parallel alignment, the magneto-resistance increases with an increasing field and for the perpendicular alignment, the magneto-resistance decreases. The effect of AMR is typically in the size of the GMR or about one magnitude smaller and with a change of the sign of A (A > 0), depending on the investigated material.

The restricted mutual solubility is known for Cu and Co. and the granular GMR was discovered on magnetron sputtered CuCo [3,4]. They found a strong dependence of the resistance with the applied magnetic field, and the resistivity being highest in the initial non-magnetized state decreases with increasing applied field. The resistivity increases again upon decreasing the field and reaches a local maximum at the coercive field. However, resistivity is slightly lower than in the initial, non-magnetized state. To give a first idea on the amount of GMR present, some results of original works [3,4] are presented: Berkowitz et al. [3] investigated Cu containing 12, 19, and 28 at% Co and found for $Cu_{81}Co_{19}$ (as an example) a GMR of 10% at 10 K at the highest applied magnetic fields of 20 kOe. Negligible GMR was found at room temperature. Xiao et al. [4] found a GMR of 16.5% at 5 K for magnetron sputtered Cu80Co20, annealed for 10 min at 500 °C. A review on GMR, including granular GMR, is provided in reference [28].

The goal of this work was to produce bulk materials of different amounts of ferromagnetic and diamagnetic components. This enables the investigation of the influence of composition and processing parameters on the evolving microstructure, on the development of small, ferromagnetic particles and thus on the GMR. The chosen method is high pressure torsion (HPT) [29], a special method of severe plastic deformation (SPD), as it provides the opportunity to easily produce bulk samples from elemental powder mixtures [30]. The principle idea of HPT is based on the work of Bridgman [31], where material is confined under high hydrostatic pressure between two anvils. One anvil is rotated against the other and the material is severely deformed by shear deformation and the microstructure gets refined. This refinement saturates at a certain grain size—mostly depending on the amount of alloying elements, impurities, and deformation temperature [32].

Regarding the investigation of different magnetic properties of materials deformed by HPT, numerous studies focusing on the magnetic properties of HPT-processed materials are available [33–42]. Other techniques to apply severe deformation onto materials include ball milling, mechanical alloying, and equal channel angular pressing (ECAP), with some studies focusing on the magnetic properties of these alternative processing routes [18–21,26,43,44]. However, to the best knowledge of the authors, there are only two studies on HPT-deformed materials, which also contain information on magneto-resistive properties [34,37]. In references [34,37], the authors used arc-melted Cu-10wt% Co for HPT deformation. The magneto-resistive drop was ~0.25% at room temperature and ~2.5% at 77 K, with both measured in fields of 6 kOe.

In summary, a detailed GMR—Study of the influences of HPT process parameters including deformation temperature and composition on GMR is lacking, which is the motivation for and aim of the presented work. Within this study, the Cu-Co-system was thoroughly investigated to understand the dependency of composition on the GMR, and to demonstrate the applicability of HPT throughout the whole compositional range.

2. Materials and Methods

Commercially available pure elemental powders were used as a starting material: Fe (MaTeck, Jülich, Germany, 99.9%, -100 + 200 mesh), Co (Goodfellow, Hamburg, Deutschland, 99.9% 50–150 µm), high purity Co (Alfa Aesar—Puratronic, Ward Hill, MA, USA, 99.998%, -22 mesh), and Cu (Alfa Aesar, Ward Hill, MA, USA, 99.9%, -170 + 400 mesh). The reason for using two grades of Co powder was the following: although the diameter of the particles of the less pure powder is rather large, the grains are small, meaning HPT deformation of high Co containing materials became difficult. For comparison, scanning electron micrographs of both types of Co powders are presented in Figure 1. Description of the other powders used is given in reference [40].

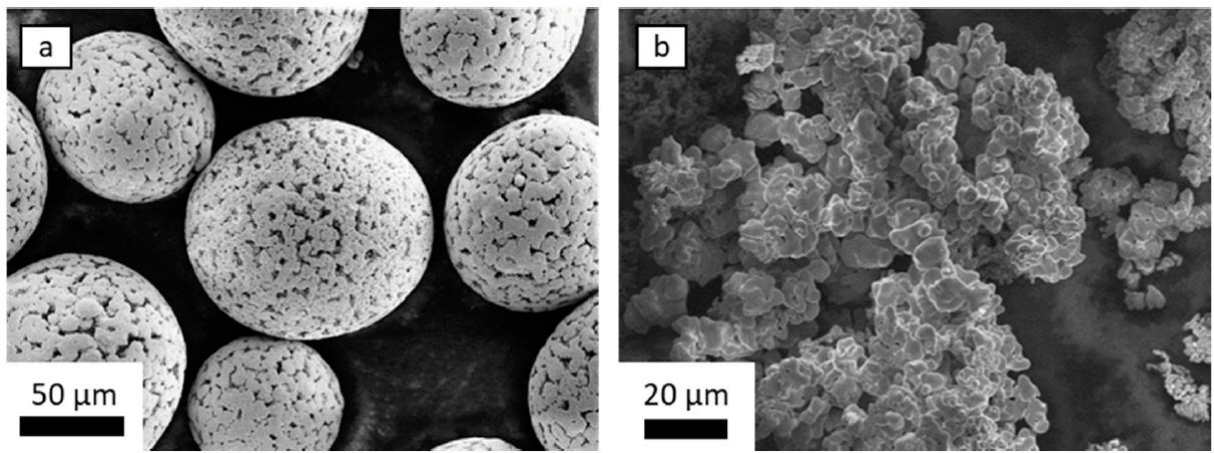

**Figure 1.** scanning electron microscopy (SEM) micrographs of Co particles. The main difference is the particle shape, the purity, and the resulting grain size. (**a**) Purity 99.9%. (**b**) Purity 99.998%.

For preventing oxidation, the powders are stored and prepared for further processing in an argon (Ar)-filled glove box. For the first processing step (pre-compaction of mixed powders in the HPT-tool), the powders are contained in a specially-made, Ar-filled compaction device [40]. After pre-compaction at a pressure of 5 GPa, the samples are deformed in air, at room temperature (RT) or at elevated temperatures, as described below (Table 1). Samples of 8 mm diameter and approximately 0.5 mm thickness were produced. To investigate the effects of varying process conditions on microstructure and the GMR, several samples of about 50/50 at% were made. They vary in sample size, processing temperature, and applied strain rate. To prove the feasibility of upscaling the HPT-process, one sample with approximately equiatomic composition was made with a different, larger HPT-tool. It is capable of applying a force of 4 MN; thus, samples of diameters of 30 mm ($p_{max}$ = 5.6 GPa) and thicknesses of several millimeters can be deformed. The strain rate which was applied to the small specimens was

~0.6 s$^{-1}$, whereas it was approximately 0.05 s$^{-1}$ for the large sample. In the latter case, the powder mixture was simply poured in the anvil's cavity and deformed under atmospheric conditions. According to Equation (2)

$$\varepsilon_{v.M.} = 2\pi n r / (t \sqrt{3}), \qquad (2)$$

the equivalent von-Mises-strain $\varepsilon_{v.M.}$ applied to the material upon HPT is given by the radius r, the number of applied rotations n, and the thickness t. For samples of 8 mm diameter, this yields an applied strain of 2200 at a radius of 3 mm after 100 rotations, assuming a thickness of 0.5 mm. For the large sample (diameter 30 mm), a strain of 1300 was applied at a radius of 10 mm (rotations: 250, thickness: ~7 mm). Although the applied strain is smaller and the strain rate is considerably lower, the large sample heats up during HPT due to plastic deformation and a lack of heat dissipation [45]. Therefore, it was deformed at elevated temperature although no external heating device was used.

**Table 1.** Summary of investigated samples, varying in composition (energy dispersive X-ray - measurement) and processing parameters including HPT-tool size, processing temperature, number of rotations, and purity of the used Co-powder (where applicable).

| Composition [at%] | HPT-Tool/Sample Diameter [mm] | Processing Temperature | No. of Rotations | Purity of Co-Powder [%] |
|---|---|---|---|---|
| Cu | Small/8 | RT | >25 | n.a. |
| $Cu_{81}Co_{19}$ | Small/8 | RT | 100 | 99.9 |
| $Cu_{64}Co_{36}$ | Small/8 | RT | 100 | 99.9 |
| $Cu_{55}Co_{45}$ | Large/30 | Elevated | 250 | 99.9 |
| $Cu_{52}Co_{48}$ | Small/8 | RT | 100 | 99.9 |
| $Cu_{49}Co_{51}$ | Small/8 | 200°C | 100 | 99.9 |
| $Cu_{43}Co_{57}$ | Small/8 | RT | 100 | 99.9 |
| $Cu_{22}Co_{78}$ | Small/8 | 300°C | 100 | 99.998 |
| Co | Small/8 | RT | 50 | 99.998 |
| $Cu_{85}Fe_{15}$ | Small/8 | RT | 100 | n.a. |

To enable a thorough microstructural and magneto-resistive investigation, at least two small HPT samples were made out of each powder mixture, but only one large sample was made since several GMR specimens and specimens for microstructural analysis could be made from one sample. All relevant sample and process parameters are summarized in Table 1.
For detailed investigations of the microstructure scanning electron microscopy (SEM) and transmission electron microscopy (TEM) studies as well as synchrotron measurements were performed. Vickers hardness was measured to ensure there was a microstructurally saturated state of the sample. GMR-specimens were taken from this microstructurally saturated region. Hardness measurements were made along the radius of the HPT-disc in a tangential direction (Micromet 5104, Buehler, Lake Bluff, IL, USA). SEM micrographs using a backscattered electron detection mode (BSE) were acquired with a LEO 1525

(Zeiss, Oberkochen, Germany) that was further equipped with an electron backscatter diffraction (EBSD) system (e⁻ Flash^FS, Bruker, Berlin, Germany) and an energy dispersive X-ray (EDX) system (XFlash 6-60, Bruker, Berlin, Germany). The conventional EBSD can easily be changed to a transmission EBSD geometry (Transmission Kikuchi Diffraction, TKD) [46,47]; thus, EBSD-TKD was performed for a chosen TEM specimen ($Cu_{55}Co_{45}$). The TEM specimen was made using conventional metallographic thinning methods and a final ion polishing step with grazing incidence. For analysing the results of EBSD-TKD and EDX measurements, the software package Esprit 2.1 from Bruker (Billerica, MA, USA) was used. The TEM/TKD—Specimen was taken at a radius of ~10 mm from the large HPT-disc, where the incident electron beam was parallel to the axial HPT-direction. Using an acceleration voltage of 30 kV and a step size of 5 nm to 9 nm (four scans in total), the orientation data was recorded. After scanning, a clean-up of the data was performed, where certain zero solutions are absorbed by the surrounding well-defined matrix. Grain boundaries were defined by having more than 15° misorientation between the two grains, and grains below the size of five pixels were rejected.

Using the same specimen, TEM micrographs were made with a JEOL-TEM (JEM2200FS, Tokyo, Japan). For phase analysis, transmissive synchrotron X-ray diffraction measurements were made at PETRA III, P07 at DESY, Hamburg, Germany (Deutsches Elektronen Synchrotron). A beam energy of 98.25 keV was used and recorded spectra were analyzed with the software PyFAI and compared with $CeO_2$–standards. Magnetic properties were measured with a superconducting quantum interference device (SQUID, Quantum Design MPMS-XL-7, Darmstadt, Germany). For small HPT-discs, the specimens for SEM and synchrotron investigations were taken out in accordance with reference [40]. For the sample made with the large HPT-tool, the TEM study was performed in an axial view. For SQUID measurements, the applied field was parallel to the axial HPT-direction.

For GMR-measurements, long and thin specimens were cut out of the HPT-disc along a secant line. According to Equation (2), the center of the HPT-disc does not receive a lot of deformation; thus, it was avoided for resistivity measurements. The closest distance of the GMR-specimen to the former center of the HPT-disc was at least 1.5 to 2 mm. The length of the GMR-specimens was several millimeters; their thickness was reduced by grinding and polishing to approximately 200 µm in order to increase the voltage drop. We sought to make sure that the impact of heat during specimen preparation was kept as low as reasonably possible; no microstructural changes are expected to occur upon sample preparation. In literature, it was found that for measurements on thin films, the applied current and magnetic field are often within the film plane [48]. Following this, the radial-tangential-plane was identified as the "film plane" for small HPT-samples. In this case, the applied current has components in the radial and shear direction. For the large HPT-sample, the current was applied in an axial direction. The specimen was placed in a four-point resistance setup, which again is placed within the air gap of an electromagnet (Type B-E 30, Bruker, Karlsruhe, Germany). The air gap was fixed with 50 mm and the diameter of the conical pole pieces was 176 mm (max. field = 22.5 kOe) providing enough space within the homogeneous magnetic field for the four-point resistance setup. The voltage was measured using a multimeter (model 2000, Keithley, Cleveland, OH, USA), the current source was a sourcemeter (model 2400, Keithley, Cleveland, OH, USA), and the applied current was 500 mA to 800 mA. The strength and stability of the magnetic field was measured using a Gaussmeter (Model 475 DSP,

Lakeshore, Westerville, OH, USA). The electromagnet's power supply, the current source, and the voltage and field measurements were connected for communication and data collection.

## 3. Results

### 3.1. Microstructural Characterization

A typical feature of HPT-deformed materials is an increasing hardness and decreasing microstructural size with an increasing distance from the sample's center. As soon as a microstructural saturation is reached, the hardness does not increase further [32]. For a valid comparison of samples of varying Co-content (Figure 2), Vickers hardness values were taken from the saturation region of constant hardness. Measurements within these regions show a clear trend of increasing hardness with increasing Co-content. Figure 2 was drawn using samples from this work and data from samples presented in reference [40]. The trend breaks down for pure Co of highest purity, whose hardness was found to be 368 HV. In the latter case, neither impurities (as would be the case for the less pure Co-powder) nor another element obstruct grain boundary motion; the grains are larger and the material becomes softer.

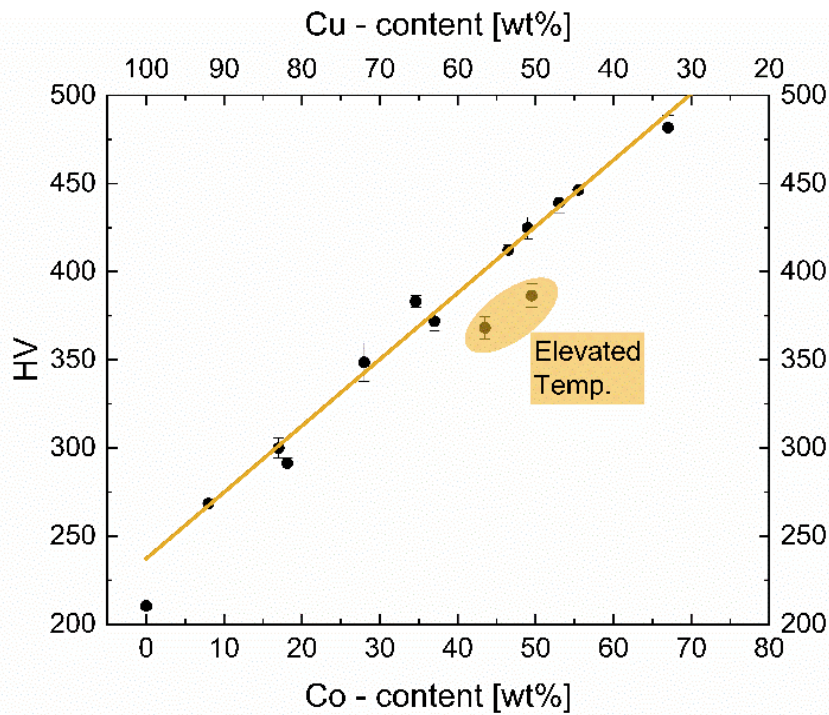

**Figure 2.** Hardness HV as a function of Cu-Co-content, showing a linear trend for room temperature deformed samples. The highlighted measurements were made on samples Cu55Co45 (large HPT, elevated temperature) and Cu49Co51 (small HPT, 200 _C) Hardness values from samples presented in reference [40] are included. The linear fit between 10 wt % and 70 wt % Co, representing a simple rule of mixture, is a guide for the eye.

SE micrographs already give a first indication of the enhanced, mutual solubility of Cu- and Co-phases upon HPT. Figure 3 shows micrographs taken in tangential direction from the saturation region for binary CuCo composites. They were made at a radius of 3 mm ($\varepsilon_{v.M}$ ~ 2200) with the exception of r = 10 mm ($\varepsilon_{v.M}$ ~ 1300) for the large sample (Figure 3c). At room temperature and for 200 °C deformed samples, the microstructure appears to be homogeneous on a scale of ~1 µm, while the microstructure for the samples deformed at elevated temperature (Figure 3c) and 300 °C (Figure 3f) appears to be granular. EDX measurements of the sample presented in Figure 3c) using low energy electrons (5 kV, smaller excitation volume) show changes in the chemical composition at the same length scale (Figure 4). The small sample $Cu_{49}Co_{51}$ (200 °C deformation temperature) does not show this segregation of Co. Therefore, the microstructure of the large sample resembles more like the one of the Co-rich Cu22Co78, which experienced a deformation temperature of 300 °C.

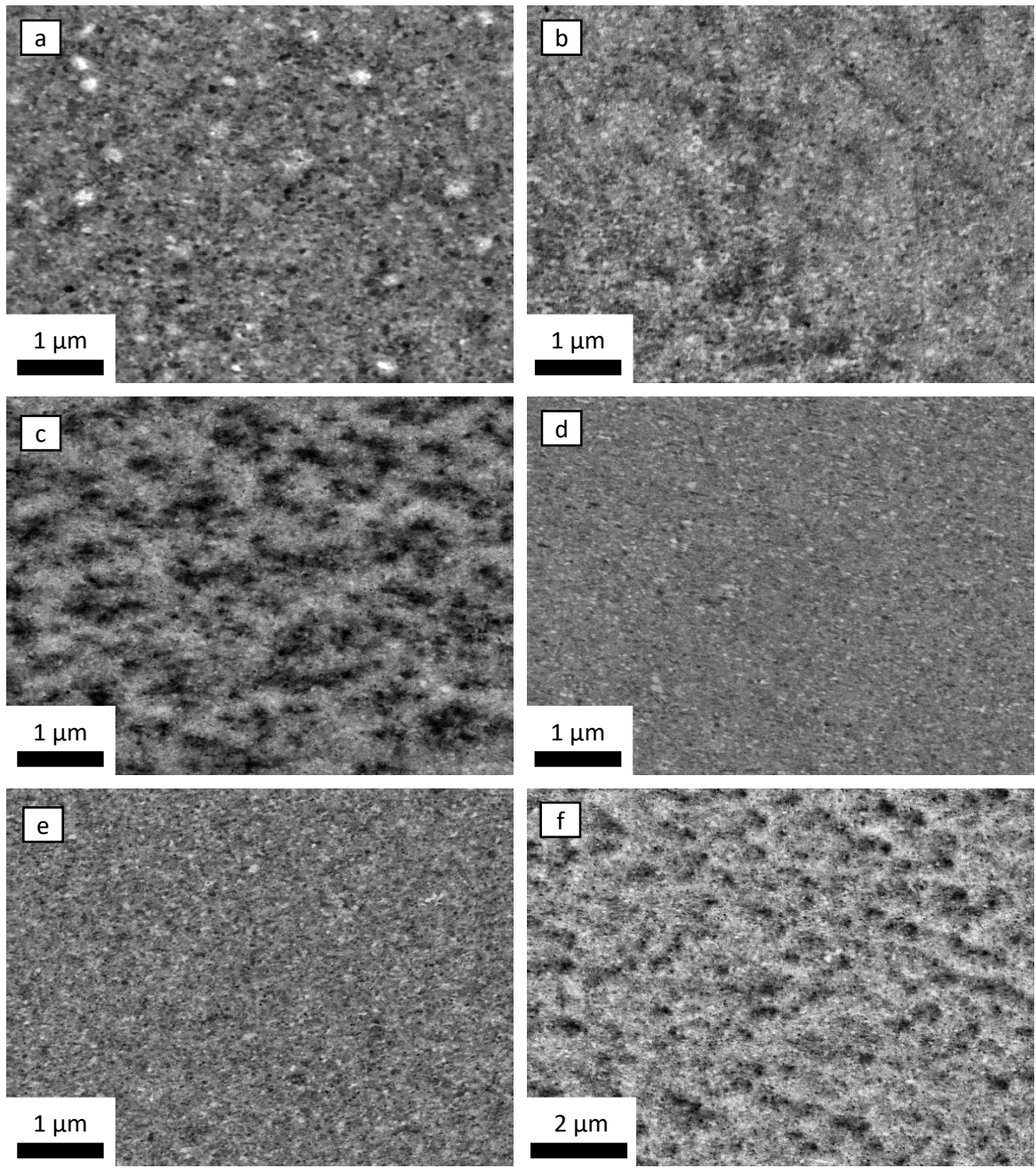

**Figure 3.** SEM micrographs of HPT-deformed samples, taken at r = 3 mm (with the exception of (**c**)) in tangential direction. (**a**) $Cu_{81}Co_{19}$; (**b**) $Cu_{64}Co_{36}$; (**c**) $Cu_{55}Co_{45}$ (r = 10 mm, large HPT-tool, deformed at elevated temperature); (**d**) $Cu_{52}Co_{48}$; (**e**) $Cu_{49}Co_{51}$ deformed at 200 °C; (**f**) $Cu_{22}Co_{78}$ deformed at 300 °C.

For the determination of the actual chemical composition and elemental distribution of the deformed samples and to detect possible deviations from the nominal composition of the powder mixture, several EDX spectra were recorded at large radii, and their mean values are presented for the chemical composition in Table 1. To demonstrate the good co-deformation and the increasing intermixing with high applied strains, 60 EDX spectra were recorded right at the center and at radii of 1, 2, and 3 mm. As an example, the sample $Cu_{85}Fe_{15}$ was used. In Figure 5a, it can be seen that still some Fe-particles (dark particles) can be found for small radii (r = 0 mm, r = 1 mm) and the distribution of results of EDX-analysis (Figure 5d) is widespread (blue lines), due to the existence of Cu-rich and Fe-rich regions. For r = 2 mm and r = 3 mm, the distribution become narrower, demonstrating improved intermixing of Cu and Fe on the scale of the EDX-analyzed volume. This is also confirmed by micrographs in Figure 5b,c.

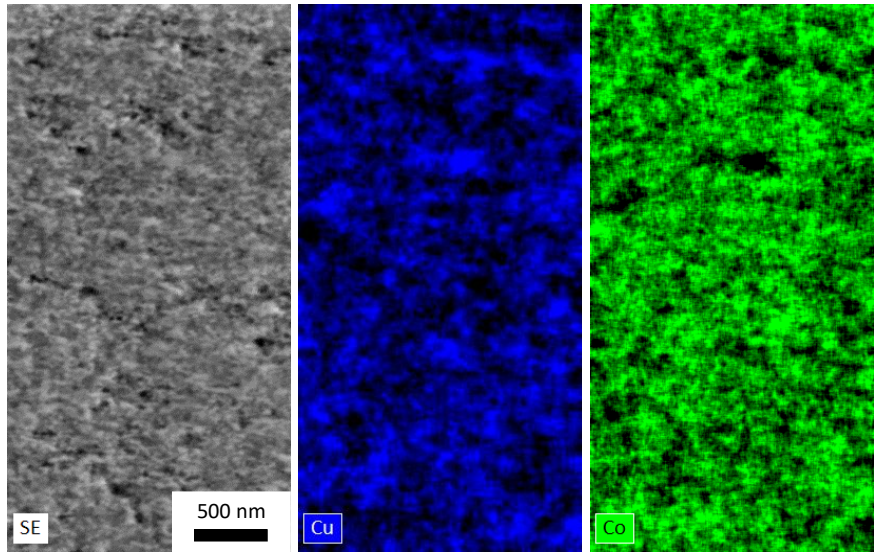

**Figure 4**. SEM micrograph of the large HPT sample $Cu_{55}Co_{45}$, taken at r = 10 mm, with EDX maps of the same region showing the distribution of Cu (center, blue) and Co (right, green).

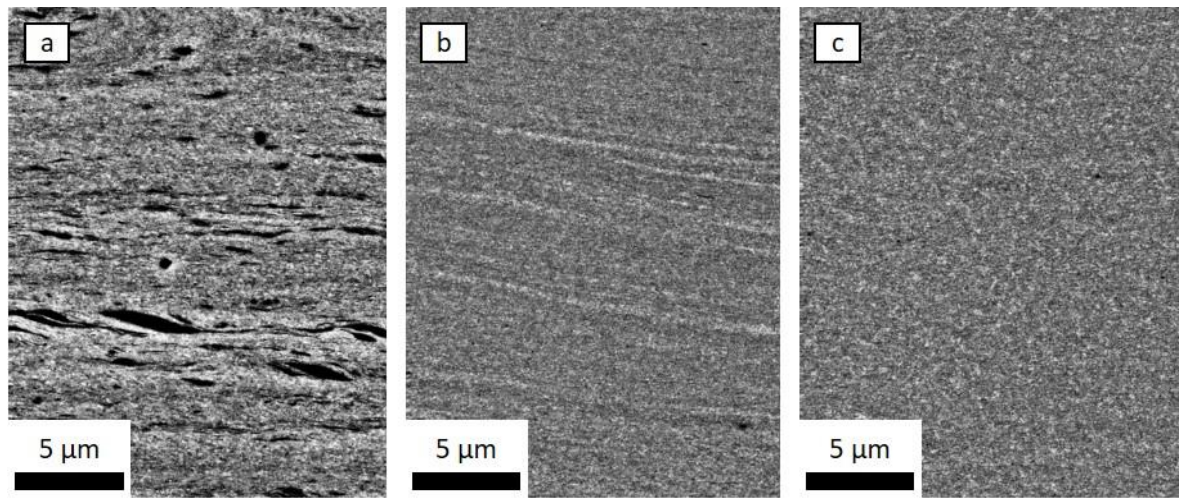

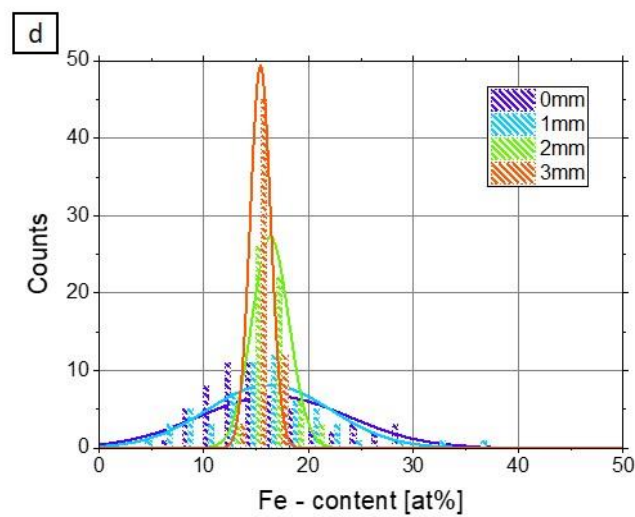

**Figure 5.** SEM micrographs of the sample containing $Cu_{85}Fe_{15}$ at (**a**) r ~ 0 mm, (**b**) r = 1 mm, (**c**) r = 2 mm, from left to right (No. rotations = 100). The progressing dissolution of Fe into the Cu matrix can be seen. (**d**) Distribution of EDX-results for the same sample and for different radii. 60 spectra were recorded at each position. As a guide for the eye, a normal distribution was fitted to the data.

As typical examples taken from different ranges of the CuCo system, the following three specimens were subjected to synchrotron diffraction phase analysis: $Cu_{81}Co_{19}$, $Cu_{55}Co_{45}$, $Cu_{22}Co\#$. The results, logarithmic intensity as a function of scattering vector q are presented in Figure 6 for different radii. To compare with, the peaks of fcc-Cu (lattice constant d = 3.615 Å [49]), fcc-Co (d = 3.554 Å [50]) and of hcp-Co, data from reference [51], are included. For the low Co-containing material, small peaks of hcp-Co can be found for r = 1 mm; however, these peaks vanish for larger radii. The Cu-peak slightly deviates from the position of pure Cu, due to supersaturating the crystal with Co. This effect is stronger for the higher-Co containing material (Figure 6b); however, traces of hcp-Co can be found for all radii. No hcp-Co can be found in the large sample Cu55Co45 (Figure 6c,d). The plateau-like shape of the peaks – especially at high q – is explained most likely due to the occurrence of two fcc-phases for Co and Cu.

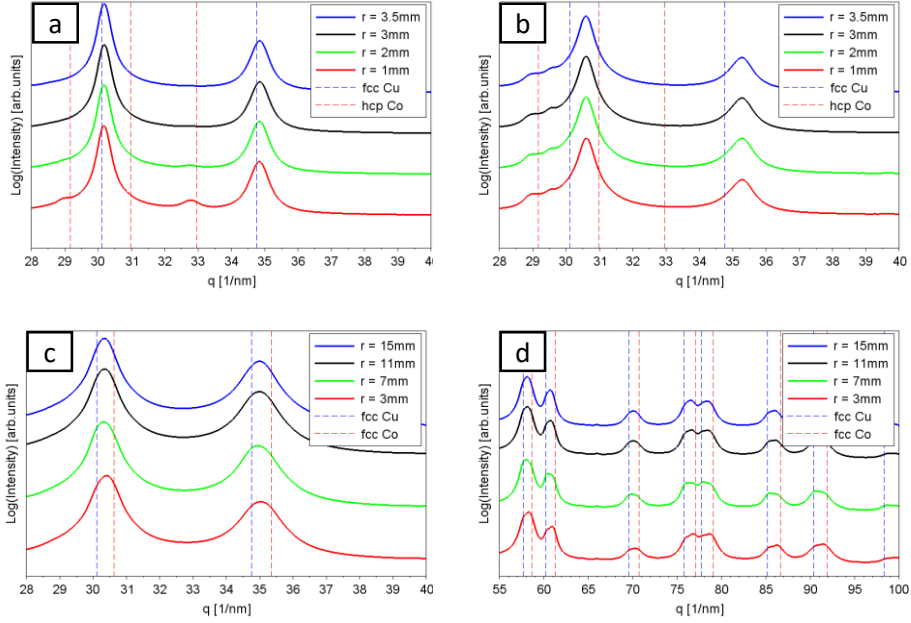

**Figure 6.** Synchrotron data for three CuCo specimens (**a**) Cu81Co19 deformed at room temperature, (**b**) Cu22Co78 deformed at 300 °C, (**c**) Cu55Co45 deformed at elevated temperature and lower applied strain rate. Note the larger radius of the specimen, (**d**) like (**c**) but showing a different regime of scattering vector q.

Further microstructural investigation, using TKD, was made for Cu55Co45 to obtained deeper insights into the as-deformed microstructure. As an example for TKD results, Figure 7a is shown. Although the hcp-Co phase was searched for, there are no clear indications of hcp-Co grains of detectable size. Distinction between grains rich in fcc-Co and rich in fcc-Cu cannot be made since the difference in lattice parameter is too small. Identified grains are in the 100 nm regime and a texture analysis, which is not shown here, yielded the shear texture typical of fcc metals [52]. In total, four scans were made, and their aggregated area weighted grain size distribution f (d) was fitted by a lognormal distribution.

$$f(d) = \frac{1}{\sqrt{2\pi}\sigma d} e^{-\frac{(\ln(d)-\mu)^2}{2\sigma^2}} \qquad (3)$$

With the mean μ and the standard deviation σ, the median value of the grain size $e^\mu$ is found to be 79 nm. Stückler et al. [40] found similar results for other CuCo materials, which were processed the same way: A decreasing grain size of 100 nm, 78 nm, and 77 nm for increasing Co-content of Cu-Co28 wt %, Cu-Co49 wt %, and Cu-Co67 wt %, respectively.

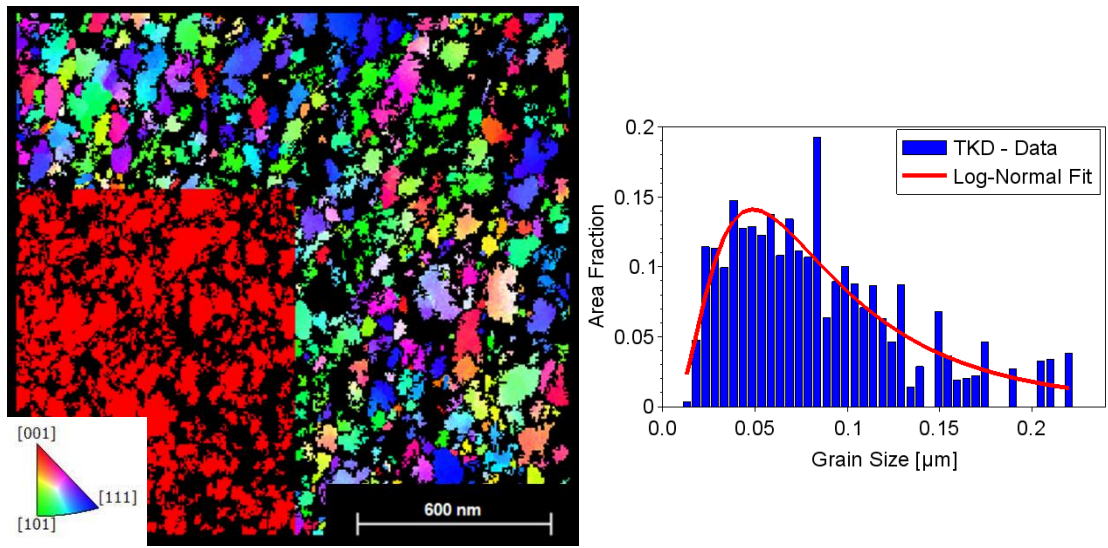

**Figure 7.** (**a**) Inverse pole figure map of $Cu_{55}Co_{45}$, taken at a radius of ~10 mm in an axial direction. As an overlay in the lower left part, the Cu-phase map (fcc–phase map, respectively) of the identical area is shown. After the clean-up, all identified grains are found to be fcc. No hcp-particles can be detected. (**b**) Aggregated grain size distribution from four TKD scans. In total, 1648 grains were taken into account.

Regarding spatial resolution, TKD-EBSD is better than conventional EBSD in many cases. Ge et al. [17] made a related analysis on Co-particle sizes in electrodeposited $Cu_{84}Co_{16}$, using TEM, before and after annealing for 30 min at 695 K and found mean values of 10 nm and 12 nm respectively, being even below the resolution limit of TKD. Thus, a high angle annular dark field (HAADF) image was recorded in scanning TEM, Figure 8. HAADF is sensitive to the atomic number, while Co-enriched (dark) and Cu-enriched (bright) regions can be identified within the matrix.

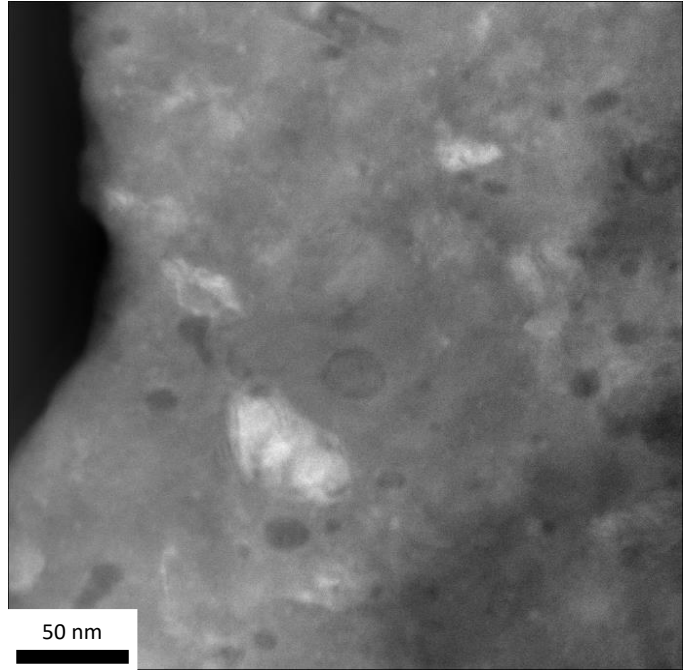

**Figure 8.** High angle annular dark field (HAADF) image of Cu$_{55}$Co$_{45}$, using the same specimen, which was used for deriving the results shown in Figure 7. The Co-particles (dark regions) with a size of several tens of nanometers become visible.

### 3.2. Magnetic Properties – Magneto-Resistivity

To investigate the change in magneto-resistive behavior for all produced samples, four-point resistance measurements were performed at room temperature in magnetic fields up to 22.5 kOe. The results of the drop in resistance with applied magnetic field, displayed according to Equation (1), are summarized in Figures 9 and 10. Two different testing directions were taken into account. First, applied current and applied magnetic field are parallel; second, current and magnetic field are perpendicular to each other. Probe current flows within the shear-tangential HPT-plane, except for Cu$_{55}$Co$_{45}$. Depending on the composition and processing parameters, different types of resistive behavior within magnetic fields can be identified.

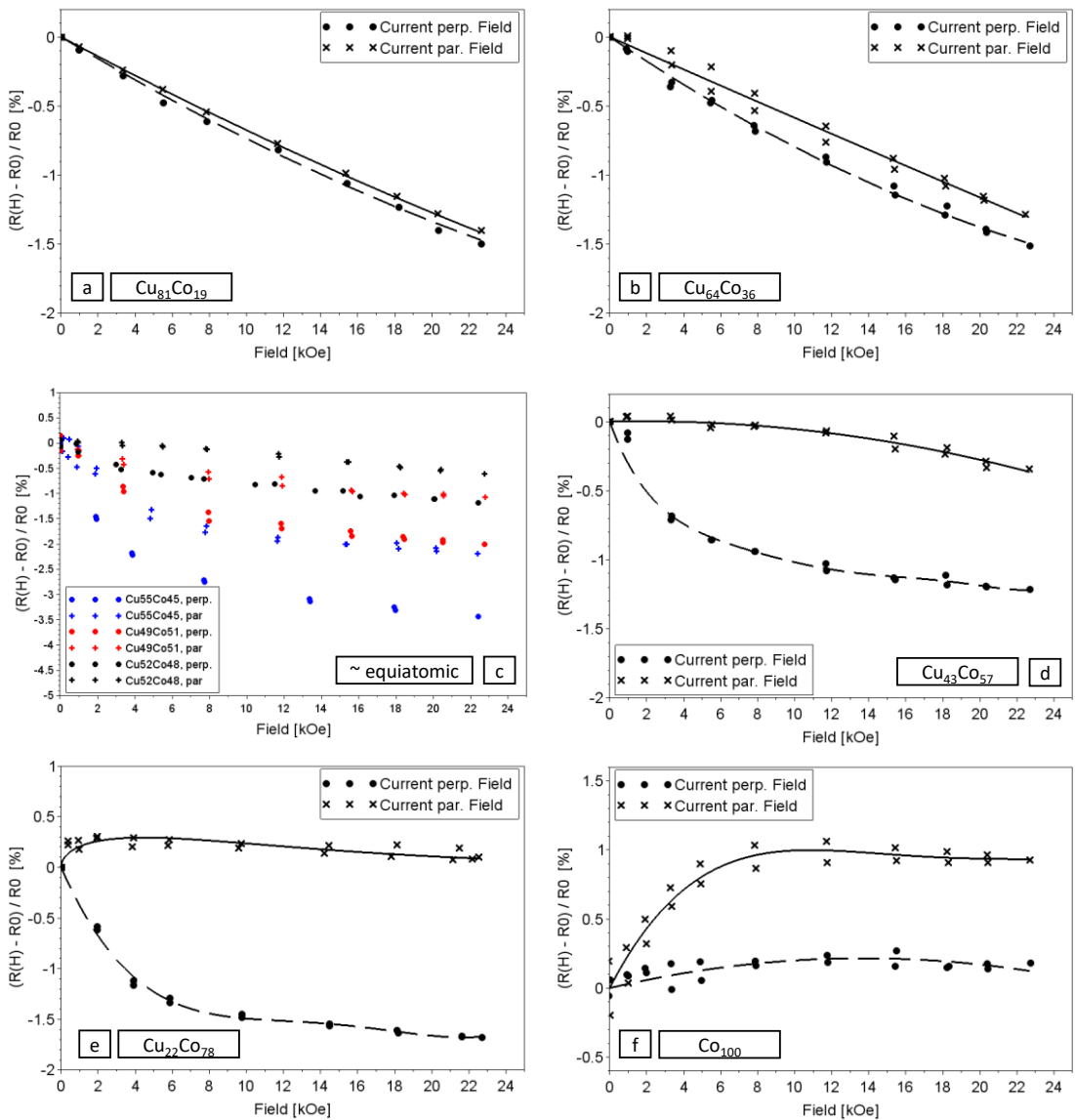

**Figure 9.** Drop in room-temperature resistance as a function of magnetic fields for Co-based HPT-deformed samples. (**a**) $Cu_{81}Co_{19}$. (**b**) $Cu_{64}Co_{36}$. (**c**) black: $Cu_{52}Co_{48}$ deformed at room temperature; red: $Cu_{49}Co_{51}$ deformed at 200 °C; blue: $Cu_{55}Co_{45}$ deformed at elevated temperature with the large HPT-tool. (**d**) $Cu_{43}Co_{57}$. (**e**) $Cu_{22}Co_{78}$, deformed at 300 °C. (**f**) Pure Co. The continuous lines (parallel alignment) and dashed lines (perpendicular alignment) are derived from polynomial fits and serve only as a guide for the eye.

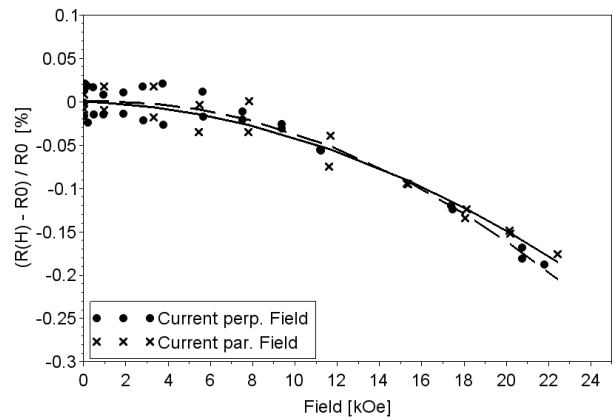

**Figure 10.** Drop in room-temperature resistance as a function of applied magnetic fields for HPT-deformed $Cu_{85}Fe_{15}$. The continuous line (parallel alignment) and dashed line (perpendicular alignment) are derived from polynomial fits and serve only as a guide for the eye.

For small Co-content, an almost isotropic dependence of resistance with magnetic field can be found (Figure 9a,b). With increasing ferromagnetic content, there is a gradual development of an anisotropic component, which is due to larger, percolating ferromagnetic regions. Thus, for medium and high ferromagnetic content, the resistance curve can be seen as a superposition of GMR and AMR. Pure Co (Figure 9f) is the best example for AMR with the resistance being higher for parallel current and magnetic field. For pure Co, the di_erence in resistivity between parallel and perpendicular orientation increases quickly for small fields; for higher fields, the curves are about parallel.

In case for replacing Co with Fe, the e_ect for low ferromagnetic content is identical for Fe and Co (Figure 10). The decrease in resistance with applied field is immediate at low fields. However, the e_ect is much smaller for Fe by around a factor of ten for fields as high as 22.5 kOe. A detailed discussion of the results of Figures 9 and 10 is provided in Section 4.2.

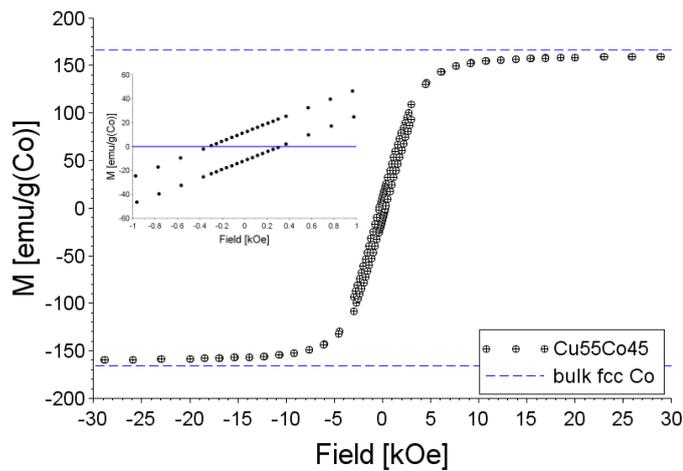

**Figure 11.** Hysteresis loop (specific magnetic moment versus applied field) of Cu55Co45, measured at 300 K, with the inset showing a detailed view for lower applied fields. For comparison, the magnetic moment of fcc-Co (164.8 emu/g [53]), is indicated as a dashed line.

The large sample deformed at elevated temperature ($Cu_{55}Co_{45}$) shows the largest reduction in resistance (~3.5% in perpendicular orientation) including a saturation of the decrease in resistance for high fields. For smaller Co content, an even larger GMR effect is expected—e.g., according to reference [17], the largest effect can be expected for a Co-concentration of 16%. Thin films, which were electrodeposited and annealed, were used for this work. However, as HPT-samples of these compositions saturate at very high fields due to paramagnetic contributions [40], the applied magnetic field is not large enough to saturate them. A sample that shows a satisfactory GMR as well as saturation in GMR at lower fields is $Cu_{55}Co_{45}$. It was chosen to be thoroughly investigated using SQUID. Figure 11 shows the result of hysteresis measurements at 300 K. The hysteresis was measured with the specimen's axial direction being parallel to the applied field. The coercivity was determined by linear interpolating between fields of +/- 1000 Oe and was found to be 316 Oe. For high applied fields, the saturation magnetization of bulk fcc-Co (164.8 emu/g [53]) is almost reached.

## 4. Discussion

### 4.1. Influence of Processing Conditions on Microstructural Evolution

Severe plastic deformation, especially at low temperatures, will promote the formation of supersaturated solid solutions. Diffusion, also taking place during the deformation process, will lead to segregation and phase formation, and as a result the resulting microstructure and elemental distribution depends on the strength of the one or the other process. As already displayed in Figure 2, the microstructure of the large sample made of $Cu_{55}Co_{45}$, resembles the one of high Co-content, deformed at high temperatures. For equiatomic composition however, there is quite a difference regarding Co agglomeration when comparing the sample processed with the large HPT and small HPT-samples. The following two paragraphs will discuss the influence of sample heating and strain rate on the evolving microstructure:

When deforming a sample with a large HPT tool, more volume is deformed and more strain energy ($\int \tau \gamma \, dV$) is transferred to heat during the whole experiment. Applied shear γ is proportional to r and integrating the infinitesimal volume V = 2π r h dr yields a $r^3$-dependency of strain energy. Using typical values: the hardness value (~350 HV) gives an approximate shear strength of 500 MPa and 150 s per rotation result in a power of ~150 W for large HPT-samples, whereas one rotation taking 45 s for the small HPT-equipment results in a heating power of approximately 10 W. The increased energy released upon deforming a sample with larger HPT-equipment is counterbalanced by the decreased strain rate and by the increased diameter of the sample and the anvils, representing the samples "cooling finger". The latter point loses importance for long-lasting experiments (~10 h in total for the $Cu_{55}Co_{45}$ sample) as the anvils heat up and their cooling capability decreases. As a result, a large sample, which is nominally deformed at room temperature, heats up more easily, leading to accelerated diffusion and segregation processes.

The second important point that has to be taken into account regarding microstructural evolution is the lower applied strain rate for the large HPT-tool. The strain rate is proportional to angular velocity divided by thickness times the radius and for the conditions described above, smaller samples were subjected to an approximately 12 times higher strain rate compared to the large sample. During HPT, the tendency of Cu and Co to segregate by diffusional processes is counter-acted by severe shear deformation. For the large HPT-sample segregation processes are more pronounced and as a result the microstructure evolves, as can be seen in Figure 12. The microstructure is shown in tangential view for radii r = 5 mm, 10 mm, 15 mm and a clear increase in the size of dark regions is visible. Thus, segregation is promoted for large radii and an increased local temperature could therefore be deduced for regions of higher strain rate. Edalati et al. determined a temperature difference of several degrees between the center and the rim of the HPT-disc. This was done by FEM simulations for a sample of 10 mm diameter [45].

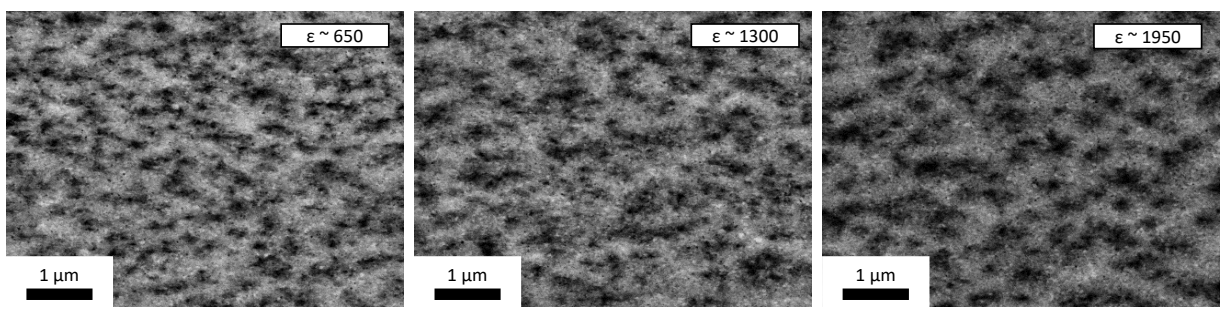

**Figure 12.** SEM—backscattered electron detection mode (BSE) micrograph of $Cu_{55}Co_{45}$, deformed with the large HPT tool at elevated temperatures. Micrographs were taken in tangential direction and for increasing radius (r = 5 mm, 10 mm, 15 mm) from left to right.

Regarding the results from synchrotron X-ray diffraction measurements (Figure 6), the low Co-containing sample ($Cu_{81}Co_{19}$) shows pure fcc phases for large radii. Diffraction lines from hcp-Co, still visible e.g., at a radius of 1 mm, vanish for higher applied strains. For high Co-contents ($Cu_{22}Co_{78}$), where deformation was possible to be performed at 300 °C, there are still hcp-Co particles remaining for large radii, as can be seen e.g., by a shoulder in the diffraction peak at q ~29 nm$^{-1}$. Nevertheless, a solid solution of the two elements can be seen for both samples. The lines are not at their nominal position for pure elements, but slightly shifted to the right (from Cu towards fcc-Co, for low-Co concentration) or to the left (from Co towards Cu, for high Co concentration). The situation is different for the $Cu_{55}Co_{45}$ sample. Here, the diffractogram is dominated by very broad peaks—especially for higher q-values, which can be easily explained by overlapping peaks at (or close to) the original position of Cu and fcc-Co.

### 4.2. Dependence of Magneto-Resistive Properties on Processing and Composition

The result of coercive field measurement from SQUID magnetometry shows a slightly increased coercivity for CuCo of about equiatomic composition, when compared to reference [40]. Therein, the powder compacted and room temperature HPT-deformed samples were measured with SQUID at 300 K and for the variety of compositions between Cu-28 wt % Co and Cu-67 wt % Co, coercive fields of ~70 Oe ranging down to ~0 Oe, below the resolution limit of SQUID magnetometry, were determined.

As described earlier, there is a variety of methods to produce thin film or bulk materials, showing a GMR. Depending on the production route, yielding different microstructures and particle size distribution—GMR is thus not only depending on the material's composition. In case the material is fully supersaturated (e.g., a minor ferromagnetic element is fully dissolved in the matrix) and no ferromagnetic particles can be found, GMR is inexistent. For precipitating ferromagnetic particles, GMR will start to rise up to a maximum value. For further growing ferromagnetic particles, GMR will decrease as the particle size are too large to form single magnetic domain particles and the particle number becomes too small to be efficient scattering sites. The dependence of the strength of GMR on the size distribution is described in reference [54].

Depending on the production route, the material might not be in the state yielding the highest GMR-effect right after processing. Very often, proper thermal treatments lead to an increase of the effect. After the production of bulk materials showing granular GMR, Ikeda et al. [20] found the maximum GMR ratio (6.4%, room temperature) for ball milled $Cu_{80}Co_{20}$ annealed at 450 °C for 1 h in vacuum. Nagamine et al. [22] report 4% (room temperature) for ball milled $(Co_{0.7}Fe_{0.3})_{20}Cu_{80}$, which was annealed for 15 min at 500 °C. For the as-milled powder they found a negligible effect (<0.2%), due to the almost perfect supersaturation of Fe and Co in Cu. Champion et al. [21] found a value of ~4% at 4.2 K for $Cu_{60}Co_{40}$ and Cu50Co50 (both in vol%) for ball milled materials. For an as-deposited, magnetron sputtered CuCo thin film, a negligible GMR at room temperature, most likely as a result of an almost perfect supersaturated state was found [3]. Comparing with our results, it is evident that the incomplete supersaturation of Cu with Co improves the GMR effect at room temperature – considering as-prepared samples. With a short annealing of the $Cu_{81}Co_{19}$ thin film [3] (10 min at 484 °C) the GMR substantially increased. Upon annealing, an increase in GMR from negligibility at room temperature to ~22% at 10 K was found.

For a granular system, the GMR might also get as large as ~50% as reported for multilayer FeCr systems at 4.2 K [1]: For another type of binary alloys, $Ag_{80}Co_{20}$ thin films, Xiong et al. [7] report on values as large as 84% (They normalize the change in resistivity to the resistance at high fields, which is different to Equation (1).) at 4.2 K for sputtered and subsequently annealed (330 °C for 10 min) specimens. Values for GMR of granular systems presented in studies on CuCo-systems are in close agreement with the values reported here. Improved thermo-mechanical treatment during HPT-processing or subsequent annealing of the as-deformed sample should lead to a closure of this small gap. Here, only HPT-processed states have been investigated and will be discussed.

In the following, the different behaviors of magneto-resistive curves shall be discussed with respect to the Co-content of the samples. Coming back to Figure 9: Starting with low ferromagnetic contents ($Cu_{81}Co_{19}$, $Cu_{64}Co_{36}$, $Cu_{85}Fe_{15}$) the dependence of the resistance with applied magnetic field is almost linear and isotropic. The effect is high for $Cu_{81}Co_{19}$ with an GMR of close to 2% for the highest applied field. No saturation in resistance drop could have been achieved even for the highest magnetic fields of 22.5 kOe. Replacing Co by Fe leads to a decrease in GMR. Through the use of SQUID magnetometry, it was shown in reference [40] that low-Fe containing CuFe composites, processed the same way as described in this work, do not saturate in magnetization even in very high fields of 70 kOe. The CuFe samples with a low Fe-content show a very pronounced paramagnetic behavior. This is explainable by a better dilution of Fe in Cu. Using GMR data from Figures 9 and 10 and comparing the GMR curves for $Cu_{85}Fe_{15}$ and $Cu_{81}Co_{19}$, which are almost identical in ferromagnetic composition, the same conclusion of an increased dilution of Fe in Cu compared to Co in Cu can be drawn.

For increasing Co content, approaching about equiatomic composition, the GMR curves for perpendicular and parallel current alignment start to differ from each other, with the curve for parallel alignment being higher. One reason could be a non-perfect globular shape of the particles. Scattering occurs at the nonmagnetic—Ferromagnetic interfaces; thus, a change in the cross-sectional area for different current flow directions could lead to a change in scattering behavior for cigar- or pancake-shaped particles. However, the specimen for probing the GMR for the large $Cu_{55}Co_{45}$ sample was not taken out the same way as for the smaller HPT-samples. At the same time, the current flow was in the tangential-radial plane (along a secant) for all small specimens, the current flow was in axial direction for the $Cu_{55}Co_{45}$ specimen. The same qualitative behavior of GMR in differently orientated specimens seems to rule out the particle shape anisotropy as an explanation for the non-perfect isotropic GMR. Another explanation for this might be the existence of large Co-particle or large percolating domains of a Co-rich phase, as these are likely to contain multiple domains. As stated in reference [10], multi-domain particles do not contribute to the GMR and as a result, a superposition of GMR and AMR occurs.

For the equiatomic composition, it was shown that deforming the sample at 200 °C leads to increased GMR compared to room temperature processing. The drop in resistivity is even higher for slightly increased process temperature and strongly reduced strain rate. Using the large HPT-tool provides temperature and time and a lower strain rate, for small Co particles to develop. According to Equation (1), an increase in GMR may as well originate from a decreased overall resistivity of the specimen – as can be expected from subsequent annealing processes but also from higher processing temperatures. The measured resistances of the specimen bear a large source for errors as the geometry is of high importance and the production of perfectly prismatic specimens is difficult. Thus, just rough values but more importantly the sequence of measured resistivities shall be given: The room temperature deformed $Cu_{52}Co_{48}$ has the highest specific resistivity (~0.52 Ω mm$^2$ m$^{-1}$), followed by $Cu_{55}Co_{45}$ (~0.18 Ω mm$^2$ m$^{-1}$) and finally $Cu_{49}Co_{51}$ (~0.13 Ω mm$^2$ m$^{-1}$). The specific resistivities were calculated, taking into account the individual specimen sizes of ~ 5 mm x 1 mm x 0.2 mm. It can be stated that the higher GMR effect for $Cu_{55}Co_{45}$ is not due to the same amount of GMR sitting atop of a smaller quantity (residual resistivity) but is truly due to a change in GMR (i.e., Co-segregation behavior), which is a consequence of changing process parameters.

SQUID measurements (Figure 11) show that the magnetization is almost fully saturated in fields of about 20 kOe and thus the approach of drawing the change in resistivity versus relative magnetization according to reference [4] can be followed. A quadratic fit of relative resistivity drop as a function of $M/M_s$ (Figure 13) yields a proportional constant A of -0.031. This value is lower than the one stated in reference [4], who used magnetron sputtered $Cu_{84}Co_{16}$, resulting in a value of -0.065 at a temperature of 5 K. The shape of the two MR-curves for this composition (Figure 9c) can be explained by a parallel connection of varistor-like components as shown in Figure 13b). On the one hand, there is conduction in the Cu phase, where small Co particles can be found. On the other hand, there is conduction in the partially percolating Co-phase, which gives rise to an anisotropic behavior. Both pathways have individual dependencies on the direction of applied magnetic field. The strength of both resistive branches determines the sign and magnitude of the proportional constant A.

With further increasing Co content, the fraction of AMR becomes more pronounced but still a markedly high drop in parallel alignment can be seen. For pure Co, the difference between both types of about 1% matches the value given by McGuire and Potter [55] of 1.9%. The difference might be explained by the overall higher resistivity of HPT-deformed, nanocrystalline Co, reducing the relative fraction of the AMR regarding total resistivity.

For pure Cu, no change in resistivity with applied magnetic field has been found within the accuracy of the used measurement setup.

The number of parameters influencing the microstructure and – as a consequence – the magneto-resistive properties is very large. Although this study is very detailed, it is not complete and leaves plenty of ideas for further investigations. In future, further interesting tasks shall be tackled: (i) Larger GMR effects will be investigated by measuring the most promising samples at cryogenic temperatures. (ii) The influence of subsequent thermal treatments of the as-HPT-deformed materials on the microstructure as well as on magneto-resistivity will be investigated.

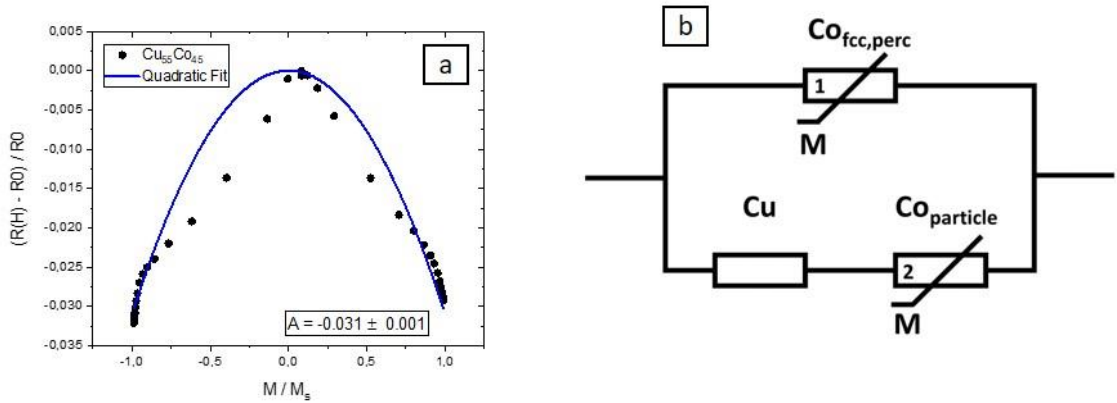

**Figure 13.** (**a**) Quadratic fit of room temperature GMR data for Cu55Co45 in accordance with reference [4]. (**b**) Schematic representation of electron conduction channels in case of Cu55Co45, of Cu, containing Co-particles on the one side and Co enriched, percolating areas on the other side.

## 5. Conclusion

The influence of the SPD conditions such as deformation temperature and strain rate on the microstructural evolution and the magneto-resistive response of different combinations of ferromagnetic and diamagnetic elements has been investigated and the following conclusions can be drawn:

It has been shown that not only the processing temperature but also the strain rate is a very important parameter regarding the deformation behavior and microstructural evolution of composite materials. The strain rate influences, in combination with the applied (or naturally evolving) temperature, the diffusion, segregation, and dissolution mechanisms taking place during severe plastic deformation. As a result, SPD by HPT is a versatile tool for achieving different microstructural states and particle sizes, respectively, when the process parameters are chosen wisely.

Depending on the ferromagnetic content of the HPT-deformed materials, different behaviors regarding magneto-resistivity at room temperature develop. When there is a small ferromagnetic content, isotropic magneto-resistive behavior (GMR) can be found. The highest drop in resistivity that could be measured within the available magnetic field was found for an approximately equiatomic composition of Cu and Co. This sample was deformed at elevated temperatures and—in respect to typical HPT-deformation processes—at a small strain rate. For medium and high Co-content, the characteristics of magneto-resistance show the occurrence of both GMR and AMR.

The investigations of GMR in connection with HPT-deformed materials are interlinked: On the one side, it is possible to first adjust the occurrence of particles and then adjust the particle size distribution. This can be done by first changing all of the material's composition and then by changing the HPT-process parameters such as deformation temperature and strain rate. On the other side, GMR-measurements are a versatile tool to study the as-deformed (and annealed) microstructures regarding the distribution of ferromagnetic particles, thereby gaining deeper insights on the deformation and segregation mechanisms acting during high pressure torsion.

**Author Contributions:** Conceptualization, S.W., R.P. and A.B.; methodology, S.W. and A.B.; software, S.W.; validation, S.W.; formal analysis, S.W.; investigation, S.W., M.S., L.W., P.K., H.K.; resources, P.K., H.K., R.P., A.B.; data curation, S.W.; writing—original draft preparation, S.W.; writing—review and editing, S.W., L.W., M.S., H.K., P.K, R.P., A.B.; visualization, S.W., M.S.; supervision, A.B.; project administration, A.B.; funding acquisition, A.B.

**Funding:** This project has received funding from the European Research Council (ERC) under the European Union's Horizon 2020 research and innovation programme (Grant No. 757333).

**Acknowledgments:** The authors are in deep gratitude to Roland Grössinger, who passed away in 2018. He always was a source of fruitful and supportive discussion and he arranged the transfer of parts of the used equipment from the Technical University of Vienna to the Erich Schmid Insitute of Materials Science. Without him, these investigations of severely deformed materials would not have been possible. The measurements leading to some of the results presented here, have been performed at PETRA III P07 at DESY Hamburg (Germany), a member of the Helmholtz Association. The authors thank of the assistance of Norbert Schell and Karoline Kormout, Stefan Zeiler, Florian Spieckermann, Pradipta Gosh and Niraj Chawake for helping with synchrotron measurements and analysis. S.W. deeply appreciates the help of Christoph Gammer for performing the TEM analysis, Mirjam Spuller and Alexander Paulischin for performing the GMR specimen preparation and assistance in resistance measurements.

**Conflicts of Interest:** The authors declare no conflict of interest.